\documentclass{emulateapj}

\newcommand{\sgra}{Sgr~A*}
\newcommand{\um}{$\mu$m}
\newcommand{\rmm}{rad~m$^{-2}$}
\newcommand{\msunyr}{$M_\sun$~yr$^{-1}$}

\newcommand{\kl}{k$\lambda$}

\begin{document}

\title{Interferometric Measurements of Variable 340~GH\lowercase{z} Linear
Polarization in Sagittarius A*}
\author{Daniel~P.~Marrone, James~M.~Moran, Jun-Hui~Zhao, 
Ramprasad~Rao\altaffilmark{1}}
\affil{Harvard-Smithsonian Center for Astrophysics, 
	60 Garden Street, Cambridge, MA 02138}
\email{dmarrone@cfa.harvard.edu}
\altaffiltext{1}{Presently with Academia Sinica Institute of Astronomy
and Astrophysics} 
\slugcomment{}
\journalinfo{To appear in The Astrophysical Journal}

\begin{abstract}
Using the Submillimeter Array, we have made the first high angular
resolution measurements of the linear polarization of Sagittarius~A*
at submillimeter wavelengths, and the first detection of intra-day
variability in its linear polarization. We detected linear
polarization at 340~GHz (880~\um) at several epochs. At the typical
resolution of 1\farcs4$\times$2\farcs2, the expected contamination
from the surrounding (partially polarized) dust emission is
negligible. We found that both the polarization fraction and position
angle are variable, with the polarization fraction dropping from 8.5\%
to 2.3\% over three days. This is the first significant measurement of
variability in the linear polarization fraction in this source. We
also found variability in the polarization and total intensity within
single nights, although the relationship between the two is not clear
from these data. The simultaneous 332 and 342~GHz position angles are
the same, setting a one-sigma rotation measure (RM) upper limit of
$7\times10^5$~\rmm.  From position angle variations and comparison of
``quiescent'' position angles observed here and at 230~GHz we infer
that the RM is a few$\times10^5$~\rmm, a factor of a few below our
direct detection limit. A generalized model of the RM produced in the
accretion flow suggests that the accretion rate at small radii must be
low, below $10^{-6}-10^{-7}$~\msunyr\ depending on the radial density
and temperature profiles, but in all cases below the gas capture rate
inferred from X-ray observations.
\end{abstract}
\shorttitle{Variable Linear Polarization in Sgr~A*}
\shortauthors{Marrone et al.}

\section{Introduction}
The radio source Sagittarius~A* (\sgra) has been conclusively
identified in the radio and infrared with a black hole of mass
$\sim3.5\times10^6 M_{\odot}$ at the center of our galaxy
\citep{ReidBrun04,SchodelE03,GhezE05,EisenhauerE05}. {\sgra} is the
nearest super-massive black hole, 100 times closer than its nearest
neighbor, M31*, and therefore should provide a unique opportunity to
understand the physics and life cycle of these objects. For a black
hole of its size, {\sgra} is extremely under-luminous, only a few
hundred solar luminosities and $10^{-8} L_{Edd}$. This surprisingly
low luminosity has motivated many theoretical and observational
efforts to understand the processes at work very near to {\sgra}.

Accretion models of {\sgra} generally seek to explain its faintness
through inefficient radiative and accretion processes. A variety of
physical mechanisms can be invoked to suppress accretion and
radiation, including convection \citep{QuatGruz00-CDAF}, jets
\citep{FalckeE93}, advection of energy stored in non-radiating ions
\citep{NaraYi94}, and winds \citep{BlandBegel99}. Many models
incorporating combinations of these and other phenomena are able to
account for the spectrum and low luminosity of {\sgra}. Therefore, the
physics of this source are not well constrained by these observations
alone.

In recent years, millimeter and submillimeter polarimetry has emerged
as an important tool for studies of \sgra. Linear polarization and its
variability can be used to understand the structure of the magnetic
field in the emission region and turbulence in the accretion flow, and
possibly to constrain the mechanisms responsible for the
multi-wavelength variability of this source. Through Faraday rotation
of the linear polarization, we can examine the density and magnetic
field distributions along the line of sight, and eventually, in
the context of more comprehensive models of the accretion flow
structure, infer an accretion rate at the inner regions of the
accretion flow \citep{QuatGruz00-LP,Agol00,MeliaLiuCoker00}.

Previous observations of the linear polarization of \sgra\ have found
low ($<$1\%) upper limits at 22, 43, and 86~GHz \citep{BowerE99-lp},
with a 2\% limit at 112~GHz \citep{BowerE01}. The lowest frequency
detection of linear polarization is at 150~GHz
\citep{AitkenE00}, suggesting that these polarimetric probes of \sgra\
can only be exploited at short millimeter and submillimeter
wavelengths. \citet{AitkenE00} found that the polarization fraction
rises steeply from 150 to 400~GHz, although these observations were
made with a single-aperture instrument and therefore required careful
removal of contaminant emission within the telescope beam. The steep
spectrum and a jump in the polarization position angle between 230 and
350~GHz in the \citet{AitkenE00} data have been taken as evidence of a
transition to optically thin synchrotron emission
\citep[e.g.,][]{AitkenE00,Agol00,MeliaLiuCoker00}.
Subsequent interferometric monitoring of the 230~GHz polarization,
with angular resolution sufficient to avoid contamination from the
surrounding emission, have shown that the 230~GHz polarization
fraction appears to remain constant over 5 years, despite variations
in the position angle on month to year timescales
\citep{BowerE03,BowerE05}. This variability reduces the significance of
the observed position angle jump and demonstrates the need for
contemporaneous measurements at multiple frequencies. \citet{BowerE05}
attribute the variations in the 230~GHz polarization to
few$\times10^5$~\rmm\ changes in the rotation measure (RM), probably
in the accretion medium, rather than to changes in the intrinsic
source polarization. As of yet, no observations have been able to
determine the RM, but they can place upper limits on the magnitude of
the RM and infer temporal variations that are within a factor of a few
of the upper limits.

Circular polarization has also been detected in this source, with a
rising polarization fraction from 1.4 to 15~GHz
\citep{BowerE99-cp,SaultMacq99,BowerE02}. Some models seeking to
explain the millimeter/submillimeter linear polarization have also
predicted appreciable circular polarization at these high frequencies
due to the conversion of linear to circular polarization in a
turbulent jet \citep{BeckertFalcke02,Beckert03}. However, measurements
to date at or above 100~GHz
\citep[e.g.,][]{TsuboiE03,BowerE03,BowerE05} have not shown circular
polarization at the percent level.

The Submillimeter Array (SMA) has the potential to contribute many new
capabilities to these studies. It provides the first opportunity to
measure the polarization above 230~GHz at angular resolution
sufficient to separate \sgra\ from its surroundings. Its large
bandwidth (2~GHz per sideband), low latitude, and dry site make it far
more sensitive for studies of this southern source than the 230~GHz
observations of \citet{BowerE03,BowerE05}, which were made with the
Berkeley-Illinois-Maryland Association array at Hat Creek,
California. Given the sensitivity and the large (10~GHz) sideband
separation, 340~GHz polarimetry with the SMA should improve limits on
the RM, and future 230~GHz polarimetry may measure it directly. These
advantages also apply to measurements of variability in total
intensity and polarization, and of circular polarization. Here we
present the first high angular resolution observations of the
submillimeter polarization of \sgra, using the newly dedicated SMA and
its polarimetry system. Our observations and reduction are discussed
in \S~\ref{s-obs}, the data and their relation to previous polarimetry
in this source in \S~\ref{s-res}, and the implications of these new
results in \S~\ref{s-disc}. We offer concluding remarks in
\S~\ref{s-concl}.

\section{Observations}
\label{s-obs}
{\sgra} was observed on several nights in 2004 using the Submillimeter
Array\footnote{The Submillimeter Array is a joint project between the
Smithsonian Astrophysical Observatory and the Academia Sinica
Institute of Astronomy and Astrophysics, and is funded by the
Smithsonian Institution and the Academia Sinica.} 
\citep{Blundell04,HoMoranLo04}. The observing dates, zenith opacity,
number of antennas used in the reduction, and on-source time are given
in Table~\ref{t-obslog}. The local oscillators were tuned to a
frequency of 336.7~GHz, centering the 2~GHz wide upper and lower
sidebands (USB and LSB) on 341.7 and 331.7~GHz, respectively. This
frequency choice avoided strong spectral lines and provided a
reasonable match to the frequency response of the SMA polarimetry
hardware, as discussed below. Our \sgra\ tracks generally included
source elevations between 20\degr\ and 41\degr (transit), a period
of seven hours, although weather, calibration, and technical problems
caused variations in the coverage. In the SMA ``Compact-North''
configuration we sampled projected baselines between 8 and
135~\kl. The average synthesized beam was approximately
1\farcs4$\times$2\farcs2. According to the estimate in
\citet{AitkenE00}, polarized emission within the 14\arcsec\
beam of the JCMT at 350~GHz contributes 100~mJy of polarized flux
density. With a beam smaller by a factor of 60, and reduced
sensitivity to large-scale emission, we expect this contaminant to be
negligible in our data.

\begin{deluxetable}{ccccccc}
\tablecolumns{7}
\tablewidth{0pt}
\tablecaption{Observing Parameters\label{t-obslog}}
\tablehead{Date & \hspace{.2cm} & $\tau_{337}$\tablenotemark{a} & \hspace{.15cm} & N$_\mathrm{ant}$ &
t$_\mathrm{int}$ (min) }
\startdata
2004 May 25             & \hspace{.2cm} & 0.16 & & 7 & 100 \\
2004 May 26             & & 0.28 & \hspace{.15cm} & 6 & 160 \\
2004 July 5\hfill\hfill & & 0.11 & & 7 & 160 \\
2004 July 6\hfill\hfill & & 0.15 & & 7 & 180 \\
2004 July 7\hfill\hfill & & 0.29 & & 6 & 170 \\
2004 July 14            & & 0.23 & & 6 & 100 \\
\enddata
\tablenotetext{a}{Mean zenith opacity at the LO frequency of 337~GHz}
\end{deluxetable}

Each SMA antenna was equipped with a single linearly polarized (LP)
feed in each of its three observing bands. Ideally, interferometric
observations of linear polarization are made with dual
circularly-polarized (CP) feeds as they separate the total intensity
(Stokes I) from the linear polarization Stokes parameters (Q and
U). For polarimetry we have converted the 340~GHz LP feeds to left-
and right-circularly polarized (LCP and RCP) feeds using positionable
quartz and sapphire quarter-wave plates. The polarization handedness
was selected by switching the angular position of the waveplate
crystal axes between two positions $\pm45^\circ$ from the polarization
angle of the receiver. Although we could only measure a single
polarization in each antenna at a given time, we sampled all four
polarized correlations (LL, LR, RL, RR) on each baseline by switching
antennas between LCP and RCP in period-16 Walsh function patterns
\citep[e.g.,][]{Rao99}. For 20-second integrations, a full cycle
required just under seven minutes. These observations were made during
the commissioning phase of the SMA polarimetry hardware; details of
this instrument can be found in Marrone (2005, in preparation).

The conversion of LP to CP was not perfect, but we calibrated the
(frequency-dependent) leakage of cross-handed polarization into each
CP state of each antenna in order to properly determine source
polarizations. We used a long observation of a polarized point source
(in this case the quasar 3C279) to simultaneously solve for the quasar
polarization and leakage terms \citep[e.g.,][]{SaultE96}. This
polarization calibration was performed twice, on May 25 and July 14,
yielding consistent leakages. The derived polarization leakages were
at or below 3\% in the USB and 5\% in the LSB, with the exception of
antenna 3, which used a sapphire waveplate with different frequency
response and poorer performance (6\% LSB leakage) than the other
waveplates. Theoretical considerations of our design suggest that the
real components of the L$\rightarrow$R and R$\rightarrow$L leakages
should be identical for a given waveplate at a given frequency, and a
comparison of the results on the two nights (a total of four
measurements of each real component) show that the rms variations in
the measured leakage terms were below 1\% for all antennas except
antenna 7. One measured leakage on July 14 was responsible for this
antenna's large rms, and because of the disagreement between the real
part of the L$\rightarrow$R and R$\rightarrow$L leakages we know that
this measurement was in error. Using the same comparison on the other
antennas we found that on average the solutions for this date were of
poorer quality, probably due to the difference in
weather. Accordingly, we adopted the May 25 leakage values for all
dates, although that required that we not use antenna 8, which was
absent from that calibration track. Errors in the leakage calibration
produce effects of varying importance, as outlined in
\citet{SaultE96}, the most important, for our purposes, are the
contamination of Q and U by Stokes I due to errors in the
determination of the leakage calibrator polarization. We have examined
this effect by comparing the Q and U fractions across sidebands on the
high signal-to-noise 3C279 data sets of May 25 and July 14; the two
sidebands should give identical measurements of Q and U from the
source, and differences can be ascribed to noise in the images and the
difference of the independent errors in the leakage solutions in the
two sidebands. With this procedure we found no inter-sideband
differences that were consistent across the two data sets, and the
differences present were consistent with the noise level, roughly
0.3\% or smaller. Because an important part of our analysis is the
comparison of position angles across sidebands, we had to ensure that
the calibration did not create a position angle offset between the
sidebands. Fortunately, although leakage errors could introduce
spurious Stokes Q or U polarization, the phase difference between the
RCP and LCP feeds, corresponding to a rotation of the sky
polarization, is identically zero because each pair of CP feeds is in
reality a single LP feed looking through both crystal axes of the same
waveplate. Therefore, the only way to create a relative position angle
difference between the sidebands would be through the leakage errors
and the resulting contamination of Q and U, an effect which appears to
be small in our data.

The flux density scale was derived from observations of Neptune on all
nights except May 25 and July 14. We expect the absolute calibration
to be accurate to about 25\% on these nights. The May 25 flux density
scale was transferred from three quasars that were also observed on
May 26; these appeared to have the same relative flux densities on
both nights to better than 10\%, consistent with the overall uncertainty
on that night, so we do not expect that the May 25 flux densities are
any more uncertain than the others. The July 14 data were obtained in
an engineering track primarily aimed at obtaining a second
polarization calibration, so only three sources are present (\sgra,
3C279, and 1743-038). Fortunately, 1743-038 has been very stable
during more than two years of monitoring observations with the SMA (an
rms flux density variation of only 20\% in that time), with even
smaller ($<10$\%) variations observed from July 5$-$7, so we have used
it as our flux density standard for the final track.

\begin{figure*}
\plotone{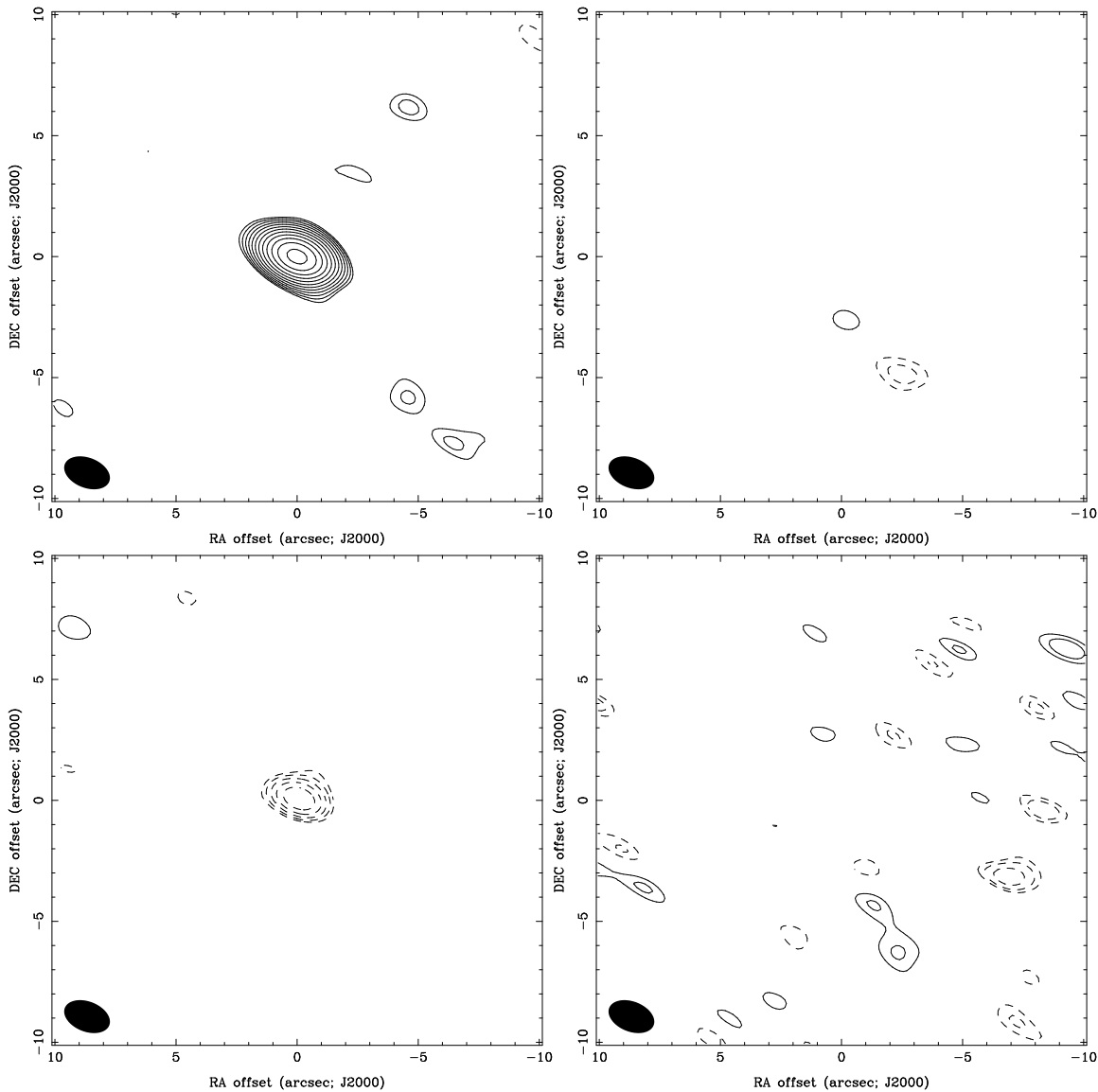}
\caption{Sample Stokes images of \sgra, from the USB data (341.7~GHz)
of 2004 May 25. The synthesized beam is
2\farcs0$\times$1\farcs2. Clockwise from top left are: I, Q, V,
U. Contours are spaced by geometrical factors of $\sqrt{2}$. For
Stokes I they are drawn at -4.2 (absent), -3, 3, 4.2, 6, 8.5, 12, 17,
24, 34, 48, 68, 96, and 136 times the 25~mJy per beam noise in the
image, for Q, U and V at -12, -8.5, -6, -4.2, -3, 3, 4.2, and 6
(absent in all three images) times the 15~mJy per beam rms noise in
the Q and U images (0.4\% of I$_{peak}$). The V contours match the Q
and U contours to highlight the increased noise introduced by the
contamination of V by I, which is due to relative gain variations
between LCP and RCP.}
\label{f-stokes}
\end{figure*}

The data were averaged over the 7 minute polarization cycle to
simulate simultaneous measurement of all four polarized visibilities,
then phase self-calibrated using the LL and RR visibilities. Quasars
were interleaved into the observations of \sgra\ to allow variability
monitoring and independent gain calibration. Transferring gains from
the quasars, rather than self-calibrating, generally resulted in
slightly lower signal-to-noise but did not change the polarization. We
attribute the increased noise ($\sim20$\%) to the 16\degr$-$40\degr\
angular separation between \sgra\ and the quasars. Following
calibration, each sideband was separately imaged in Stokes I, Q, U,
and V, using only baselines longer than 20~\kl, and then
cleaned. Sample Stokes images are shown in Figure~\ref{f-stokes}. On
July 14, due to poorer coverage in the {\it uv} plane in the short
track, we increased the cut to 30~\kl. Flux densities were extracted
from the center pixel of each image, and these are listed in
Table~\ref{t-pol}. We also examined the polarization by fitting point
sources to the central parts of the images; the point source flux
densities matched well with those obtained from the central pixel when
the signal was well above the noise, but the point source positions
and peak flux densities became erratic for low signal-to-noise images
(most Stokes Q and V images). Table~\ref{t-pol} also includes the
polarization fraction ($m$), which has been corrected for the
noise bias \citep[through quadrature subtraction of a $1\sigma$ noise
contribution, e.g.,][]{WardleKronberg74}, and the electric vector
position angle ($\chi$, determined as
$2\chi=\mathrm{tan}^{-1}\frac{\mathrm{U}}{\mathrm{Q}}$).

\begin{deluxetable*}{ccccccc}
\tiny
\tablecolumns{7}
\tablewidth{0pt}
\tablecaption{340~GHz Polarization Measurements of {\sgra}\label{t-pol}}
\tablehead{Date & I & Q & U & V & $m$ & $\chi$ \\
 & (Jy) & (mJy) & (mJy) & (mJy) & (\%) & (degrees)}
\startdata
2004 May 25\hfill\hfill & & & & & & \\ 
\hfill USB & 3.79 $\pm$ 0.03 &   9 $\pm$  15 & -244 $\pm$  15 &  -5 $\pm$  22 & 6.43 $\pm$ 0.39 & 136.1 $\pm$\hfill  1.7 \\ 
\hfill LSB & 3.79 $\pm$ 0.02 &  13 $\pm$  17 & -201 $\pm$  17 &  -9 $\pm$  21 & 5.28 $\pm$ 0.45 & 136.8 $\pm$\hfill  2.4 \\ 
\hfill Both & 3.79 $\pm$ 0.02 &  13 $\pm$  11 & -230 $\pm$  11 &  -5 $\pm$  17 & 6.07 $\pm$ 0.28 & 136.7 $\pm$\hfill  1.3 \\ 
2004 May 26\hfill\hfill & & & & & & \\ 
\hfill USB & 3.19 $\pm$ 0.03 & 145 $\pm$  20 & -97 $\pm$  20 & -14 $\pm$  21 & 5.43 $\pm$ 0.63 & 163.0 $\pm$\hfill  3.3 \\ 
\hfill LSB & 3.11 $\pm$ 0.02 & 104 $\pm$  18 & -138 $\pm$  18 & -10 $\pm$  22 & 5.53 $\pm$ 0.58 & 153.5 $\pm$\hfill  3.0 \\ 
\hfill Both & 3.16 $\pm$ 0.02 & 118 $\pm$  13 & -138 $\pm$  13 & -17 $\pm$  19 & 5.75 $\pm$ 0.43 & 155.3 $\pm$\hfill  2.1 \\ 
2004 July 5\hfill\hfill & & & & & & \\ 
\hfill USB & 3.23 $\pm$ 0.04 &  42 $\pm$  14 & -267 $\pm$  14 & -37 $\pm$  17 & 8.35 $\pm$ 0.44 & 139.5 $\pm$\hfill  1.5 \\ 
\hfill LSB & 3.13 $\pm$ 0.02 &  41 $\pm$  12 & -273 $\pm$  12 & -19 $\pm$  17 & 8.84 $\pm$ 0.38 & 139.3 $\pm$\hfill  1.2 \\ 
\hfill Both & 3.20 $\pm$ 0.02 &  42 $\pm$  10 & -270 $\pm$  10 & -38 $\pm$  13 & 8.52 $\pm$ 0.31 & 139.5 $\pm$\hfill  1.0 \\ 
2004 July 6\hfill\hfill & & & & & & \\ 
\hfill USB & 3.19 $\pm$ 0.02 &  58 $\pm$  21 & -169 $\pm$  21 & -15 $\pm$  25 & 5.56 $\pm$ 0.65 & 144.4 $\pm$\hfill  3.3 \\ 
\hfill LSB & 3.15 $\pm$ 0.03 &  29 $\pm$  18 & -164 $\pm$  18 & -16 $\pm$  22 & 5.27 $\pm$ 0.56 & 140.1 $\pm$\hfill  3.0 \\ 
\hfill Both & 3.18 $\pm$ 0.02 &  52 $\pm$  15 & -177 $\pm$  15 & -18 $\pm$  19 & 5.78 $\pm$ 0.49 & 143.2 $\pm$\hfill  2.4 \\ 
2004 July 7\hfill\hfill & & & & & & \\ 
\hfill USB & 2.71 $\pm$ 0.03 &  38 $\pm$  22 & -35 $\pm$  22 &  -8 $\pm$  29 & 1.72 $\pm$ 0.82 & 158.8 $\pm$ 13.7 \\ 
\hfill LSB & 2.78 $\pm$ 0.04 &  31 $\pm$  22 & -67 $\pm$  22 & -13 $\pm$  38 & 2.53 $\pm$ 0.80 & 147.6 $\pm$\hfill  9.0 \\ 
\hfill Both & 2.75 $\pm$ 0.03 &  44 $\pm$  17 & -49 $\pm$  17 & -17 $\pm$  25 & 2.32 $\pm$ 0.61 & 156.1 $\pm$\hfill  7.5 \\ 
2004 July 14\hfill\hfill & & & & & & \\ 
\hfill USB & 3.00 $\pm$ 0.03 &  37 $\pm$  27 & -243 $\pm$  27 &  14 $\pm$  32 & 8.14 $\pm$ 0.91 & 139.3 $\pm$\hfill  3.2 \\ 
\hfill LSB & 3.00 $\pm$ 0.03 &  29 $\pm$  19 & -175 $\pm$  19 & -17 $\pm$  25 & 5.87 $\pm$ 0.64 & 139.7 $\pm$\hfill  3.1 \\ 
\hfill Both & 3.02 $\pm$ 0.03 &  75 $\pm$  16 & -236 $\pm$  16 & -15 $\pm$  24 & 8.17 $\pm$ 0.55 & 143.8 $\pm$\hfill  1.9 \\ 
All days\hfill\hfill & & & & & & \\ 
\hfill USB & 3.33 $\pm$ 0.02 &  57 $\pm$  10 & -197 $\pm$  10 &  -9 $\pm$  15 & 6.15 $\pm$ 0.29 & 143.1 $\pm$\hfill  1.3 \\ 
\hfill LSB & 3.29 $\pm$ 0.02 &  49 $\pm$  10 & -202 $\pm$  10 &  -8 $\pm$  13 & 6.32 $\pm$ 0.29 & 141.8 $\pm$\hfill  1.3 \\ 
\hfill Both & 3.31 $\pm$ 0.02 &  59 $\pm$   7 & -204 $\pm$   7 & -17 $\pm$  11 & 6.39 $\pm$ 0.23 & 143.1 $\pm$\hfill  1.0 \\ 
\enddata
\tablecomments{Errors in the flux density columns are from the image
rms only, they do not include the 25\% absolute calibration
uncertainty, which applies equally to all flux densities and does not
affect the $m$ or $\chi$ columns.}
\normalsize
\end{deluxetable*}

\section{Results}
\label{s-res}
\subsection{Linear Polarization}
\begin{figure*}
\plotone{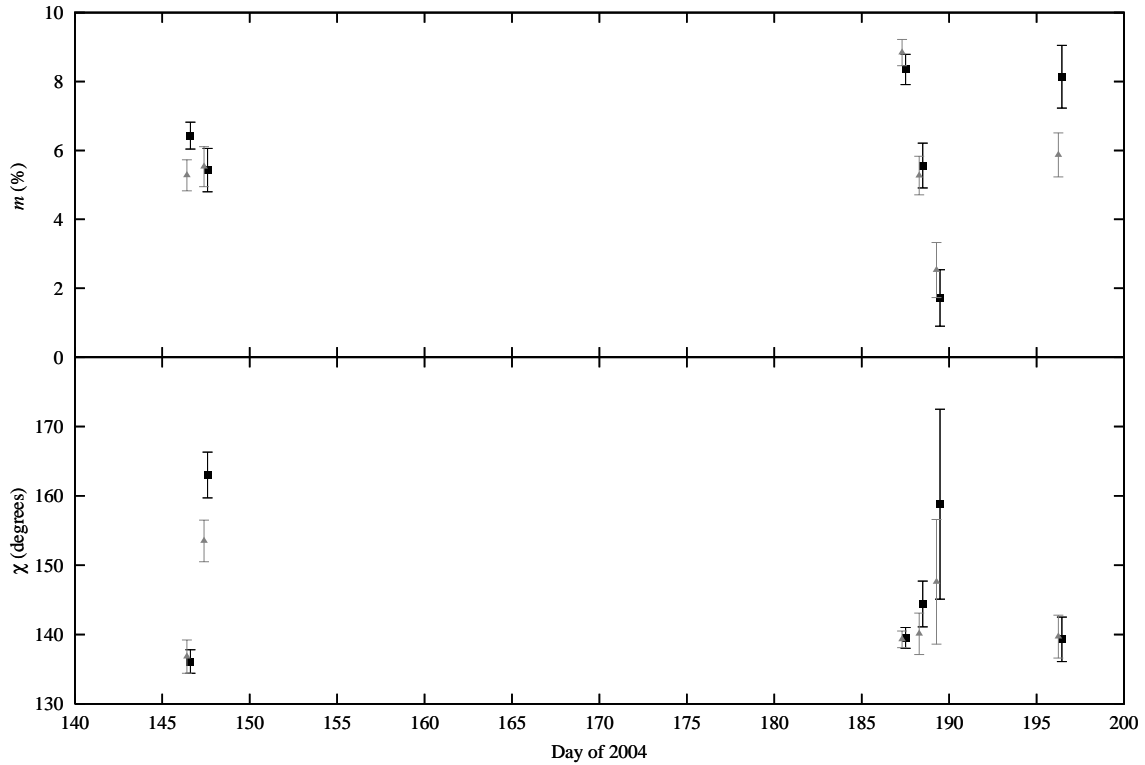}
\caption{340~GHz \sgra\ polarization fraction ($m$, {\it upper}) and
position angle ($\chi$, {\it lower}). The USB (black squares) and LSB
(gray triangles) are plotted separately for each night. The two
sidebands are slightly offset in time for clarity, but both sample the
same time interval. The large $\chi$ error bars on day 189 (July 7)
are due to the low polarization signal on that night.}
\label{f-pol}
\end{figure*}

The polarization fraction and position angle for each sideband on each
night are plotted in Figure~\ref{f-pol}. It can be seen from the
figure and the data in Table~\ref{t-pol} that we have clear detections
of the linear polarization in both sidebands on all nights. Among the
six nights of our observations, July 7 stands out for its low
polarization fraction, around 2\%. The polarization was only detected
at the $2-3\sigma$ level in each sideband, so the polarization
position angle was poorly constrained. This is the lowest linear
polarization fraction measured at or above 150~GHz, the lowest
frequency where polarization has been detected. The weather on this
night was the poorest of all the tracks, but only marginally worse
than May 26 which did not show an unusually low polarization. Other
sources in the July 7 track with measurable polarization, such as
3C279, did not show a significantly lower polarization than on other
nights, as one might have expected from a systematic problem in that
track. An obvious systematic error would be a substantial change in
the leakages with respect to previous nights; this would most easily
be caused by large changes to the alignment of the polarization
hardware. However, the hardware was not moved between installation on
July 5 and removal after the July 7 track, and the July 5 and 6 tracks
show substantially larger polarization, so this possibility seems very
unlikely. Moreover, because the leakages measured on July 14 are
consistent with the May 25 leakages, as discussed in \S~\ref{s-obs},
any change between July 6 and 7 would have to have been reversed when
the hardware was reinstalled on July 14. This low polarization
fraction, along with the unusually high polarization two nights
before, clearly demonstrates that the polarization fraction is
variable. Moreover, the polarization variations are present both in
the polarization fraction and the polarized flux density, even after
accounting for the 25\% uncertainty in the overall flux density scale,
and are not merely the result of a constant polarized emission
component with a changing total intensity.

Variability was also observed in the polarization position angle.
Polarization over four of the nights ranged between roughly 137\degr\
and 143\degr, at a weighted average position angle of 139.6\degr. The
position angle determined for May 26 differed significantly from this
range and July 7 had an extremely uncertain position angle due to the
very low polarization fraction. Neither the combined six-night data
set, nor the individual nights showed significant inter-sideband
differences, with the possible exception of May 26. On that night
$\chi_{LSB}-\chi_{USB} = (153.5^\circ\pm3.0^\circ) -
(163.0^\circ\pm3.3^\circ) = -9.5^\circ\pm4.5^\circ$, which is
marginally significant for the quoted errors. As we discussed in
\S~\ref{s-obs}, although it is possible for Stokes I to contaminate Q
and U (which determine $\chi$), this appears to be unimportant in
these data.  The 0.3\% limit on this effect is smaller than the Q and
U errors on May 26, which are 0.6\% of Stokes I. Furthermore, any
other systematic source of inter-sideband position angle offsets would
show up equally on all nights, but the six-night average
$\chi_{LSB}-\chi_{USB}$ is 1.3\degr$\pm$1.8\degr, consistent with
zero. The May 26 result is considered further in the context of a
Faraday rotation measure in \S~\ref{s-discRM}.

\subsection{Circular Polarization}
Neither the averaged data nor the individual nights show CP at a level
that is significant. The greatest deviation from zero is
$-38\pm13$~mJy on July 5, corresponding to $-1.2\pm0.4$\%. However, in
addition to the quoted error, which is the measured noise in the
cleaned map, there are well known systematic effects. The MIRIAD
reduction package \citep{miriad} uses linearized equations when
solving for the polarized leakages, ignoring second order terms in the
leakages ($d$) and linear polarization fraction. These terms
contribute a systematic error in Stokes V of the form $\mathrm{I}d^2$
and $md$ \citep{RobertsE94}, which may be of the order of a few tenths
of a percent for our leakages and the polarization of
\sgra. Moreover, the small difference in the sample times of the LL
and RR correlations on a given baseline permit gain differences, due
to weather, pointing, and system changes, to introduce differences
between the LL and RR visibilities that would not be present if these
were actually measured simultaneously (as our reduction
assumes). These gain variations contaminate Stokes V with Stokes I and
make the value of V at the peak of the I map more uncertain than the
map rms would indicate. The average of all six tracks shows
$-0.5\pm0.3$\% CP, consistent with zero, with additional systematic
error of perhaps another 0.3\%. The 0.5\% sum of these errors can be
taken as a limit on any persistent level of CP across the six nights,
and is the most stringent limit yet on CP in \sgra\ above 90~GHz.

\subsection{Intra-day Variability}
\begin{figure*}
\plotone{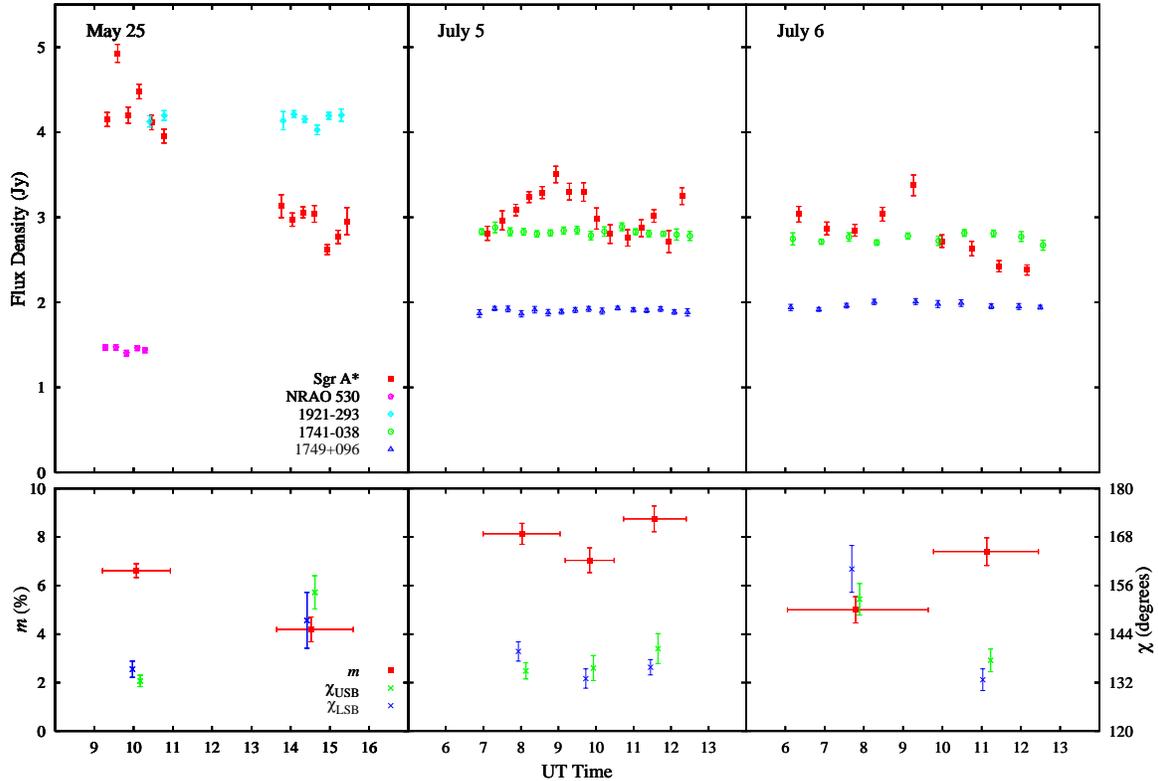}
\caption{Variability in the total intensity ({\it upper}) and
polarization (fraction and position angle, {\it lower}) of
\sgra\ at 340~GHz. The three nights with the best weather are shown,
as these permit the most accurate determinations of the polarization
variation. The total intensity light curves of the quasar calibrators,
1741$-$038, 1749$+$096, and 1921$-$293, are also shown. In the lower
plots, the binned intervals are demarcated by the horizontal bars on
the polarization fraction points. This polarization fraction is the
double-sideband value. The USB and LSB position angles sample the same
time bin but have been offset slightly from the bin center for
clarity.}
\label{f-plc}
\end{figure*}

Intra-day variability in the total intensity (Stokes I), the
polarization fraction, and position angle are shown in
Figure~\ref{f-plc}. The July 5$-$7 observations were obtained as part
of a coordinated multi-wavelength \sgra\ monitoring program, and the
observed temporal variability in Stokes I on these nights is discussed
in conjunction with results at other wavelengths in
\citet{EckartE05}. In order to prevent antennas with variable
performance from falsely modulating Stokes I, we use only the 5
antennas with the best gain stability for these light curves. Slow
variations in the gain of the other antennas are likely due to
pointing errors. We have reduced the effects of changing spatial
sampling of extended emission by removing the two baselines that
project to less than 24~\kl\ (angular scales $>$9\arcsec) during the
\sgra\ observations. Further details of the light curve reduction can
be found in \citet{EckartE05}. The variability in the linear
polarization is much harder to measure; with signals one to two orders
of magnitude weaker than Stokes I it is difficult to obtain reliable
results from a subdivided track, and we could not be as selective
about which data to exclude in the hope of removing the imprint of
instrumental variations from the polarization variation. Accordingly,
polarized light curves could not be reliably extracted for May 26 and
July 14 due to poor weather, nor July 7, due to both weather and very
low polarization fraction. The remaining three nights have been
subdivided into two or three segments at boundaries in the Stokes I
curves and the polarization has been extracted as described in
\S~\ref{s-obs}. The large (160 minute) gap on May 25, due to
instrument difficulties, served as one of the boundaries.

A great deal of variability is visible in the Stokes I curve on all
three nights, with the most notable feature being the $\sim1.5$~Jy
difference between the flux densities of the first and second halves
of the May 25 data. No such difference shows up in the light curve of
the calibrator, 1921$-$293, a source at nearly identical declination,
suggesting that this result is not an instrumental artifact. Clear
polarization variability is also measured on May 25 and July 6 in both
$m$ and $\chi$. At all times the position angles in the USB and
LSB are found to be very similar, as was observed in the full track
averages reported above.

\section{Discussion}
\label{s-disc}

\subsection{Rotation Measure}
\label{s-discRM}
The rotation measure associated with a plasma screen located between
the source and observer can be inferred from the measurement of $\chi$
at two frequencies, since it introduces a frequency dependent change
in the position angle given by
\begin{equation}
\chi(\nu) = \chi_0 + \frac{c^2}{\nu^2}\mathrm{RM} ,
\label{e-chi}
\end{equation}
where the RM is given by
\citep[e.g.,][]{GardnerWhiteoak66}
\begin{equation}
\mathrm{RM} = 8.1\times10^5 \int n_e \textit{\textbf{B}} 
\cdot d\textit{\textbf{l}}
\label{e-RM}
\end{equation}
for electron density $n_e$ in cm$^{-3}$, path length
$d\textit{\textbf{l}}$ in parsecs, and magnetic field
\textit{\textbf{B}} in Gauss. The greatest obstacle to such a
detection, as previously noted, is the variability in the
polarization, which may prevent polarization measured at different
times from being reliably compared.

The best method for measuring the RM from our data comes from the
observed difference in the simultaneous position angles in the USB and
LSB. Applying equation (\ref{e-chi}) to the two sideband frequencies
of these observations, and for position angles in degrees, we obtain
\begin{equation}
\mathrm{RM} = 3.7\times10^5 \left(\chi_{LSB}-\chi_{USB}\right) .
\label{e-RMa}
\end{equation}
Equation (\ref{e-RMa}) implicitly assumes that the Faraday rotation
occurs outside of the plasma responsible for the polarized
emission. This assumption seems reasonable for \sgra, VLBI
measurements \citep{KrichbaumE98,ShenE05,BowerE04} suggest intrinsic
sizes of $13-24 r_S$ at 215, 86, and 43~GHz, and for reasons described
in \S~\ref{s-mdot} we expect little contribution to the RM inside $300
r_S$. One other potential complication arises if the source
polarization changes with radius and the two frequencies being
compared probe different radii. For our 3\% sideband separation, and
assuming that the polarized submillimeter emission is thermal
synchrotron \citep[as is expected in ADAF models;][]{YuanE03}, we
expect a 5\% opacity difference between our sidebands, while for
non-thermal synchrotron \citep[taking an electron energy spectral
index of 2-3.5, e.g.,][]{MarkoffE01,YuanE03} the difference is
9-12\%. Emission will be contributed from a range of radii around the
$\tau=1$ surfaces, so we would have to postulate a large gradient in
the source polarization to produce a large intrinsic inter-sideband
polarization difference over such a small frequency range. Finally,
the 2~GHz bandwidth at 340~GHz limits the allowed RM to approximately
$2\times10^7$~\rmm\ if polarization is detected, as this RM would
rotate the polarization by more than a radian across the band and wash
out the signal (bandwidth depolarization). For highly polarized
emission the vector average of the polarization may still be
detectible but the position angles of the two sidebands are very
unlikely to agree in this case. We can therefore ignore the
possibility of full 180\degr\ wraps between sidebands, as a wrap
requires a RM of $7\times10^7$~\rmm.

As is clear from Table~\ref{t-pol}, we do not see a significant change
in the position angle between the two SMA sidebands on most of the
observing nights (disregarding the uncertain position angle of July
7). In the most sensitive track, July 5, the sideband difference
places a one-sigma limit of $7.1\times10^5$~\rmm\ on the RM on that
particular night, which is the most sensitive limit to date from
simultaneous interferometric observations. If the full data set is
considered together (i.e., with Stokes images derived from the
ensemble of data), the limit drops by a small amount to
$6.8\times10^5$~\rmm, although if the RM is varying between
observations this average will not actually represent a measurement of
a RM. It should be noted here that the broadband observations of
\citet{AitkenE00} were able to place a similar limit of approximately
$5\times10^5$~\rmm\ on the RM in August 1999 because of the large
bandwidth of their 150~GHz bolometer.

The May 26 sideband difference of $-9.5^\circ\pm4.5^\circ$ is possibly
significant, with an inferred rotation measure of
$(-3.5\pm1.7)\times10^6$~\rmm. If this RM had been present on the
previous night it would have shown up as a similarly large sideband
difference, instead of the observed $0.7^\circ\pm3.1^\circ$,
corresponding to a RM of $(+0.3\pm1.1)\times10^6$~\rmm. We can check
the large RM by comparing the position angles on May 25 and 26, on the
assumption that the emitted polarization ($\chi_0$ from
eq. [\ref{e-chi}]) is constant over timescales of a few days and
observed position angle changes are due to RM changes. At this
frequency, the relationship between the position angle change
($\Delta\chi$, in degrees) and the RM change is (see
eq. [\ref{e-chi}])
\begin{equation}
\Delta\mathrm{RM} = 2.2\times10^4 \Delta\chi .
\label{e-RMb}
\end{equation}
We observed an increase in the position angle from May 25 to May 26 of
$18.6^\circ\pm2.5^\circ$. If this is not a change in the
intrinsic polarization, it corresponds to an increase in the RM of
$4\times10^5$~\rmm, inconsistent with the small sideband difference on
May 25 and large difference on May 26. The position angle is 180\degr\
degenerate, however, and a $\chi$ change of
$18.6^\circ-180^\circ=-161.4^\circ$ requires a RM change of
$-3.6\times10^6$~\rmm, which agrees well with the RM inferred from the
May 26 sideband difference. It is therefore possible that we have
observed a large change in the RM between these two nights, with the May
26 value far in excess of the limits on the other five nights. We
discuss this further in \S~\ref{s-mdot}.

In the existing polarization data at 230 and 340~GHz the position
angle seems to frequently return to the same value. The
\citet{BowerE05} 230~GHz data are clustered around 111\degr\ between
2002 October and 2004 January, while four of our observations at
340~GHz have a mean position angle of 140\degr. Assuming that these
two angles sample the same $\chi_0$ (no source polarization changes
between the two observing periods or observing frequencies), we can
infer a ``quiescent'' RM of $-5.1\times10^5$~\rmm. This is just below
the RM upper limit from our most sensitive night. If the idea of a
quiescent RM is correct, then the change in the mean 230~GHz position
angle observed between early 2002 \citep{BowerE03} and 2003
\citep{BowerE05} merely reflects a change in this RM by
$-3\times10^5$~\rmm. This implies that the quiescent RM in early 2002
was around $-8\times10^5$~\rmm, which is conveniently below the
detection limit of the \citet{BowerE03} observations. If this scenario
is correct, the RM should be detectable by the SMA at 230~GHz, where
it would be observable as a 5\degr\ sideband difference.

\subsection{Accretion Rate Constraints}
\label{s-mdot}
Much of the importance placed on the RM determination stems from its
use as a probe of the accretion rate near the black hole. However, the
interpretation of a RM detection, or limit, in terms of an accretion
rate requires a model for the density and magnetic field in the
accretion flow, as these quantities actually determine the RM through
equation~(\ref{e-RM}). 

To estimate the RM predicted for a variety of accretion models we make
several simplifying assumptions. First, we assume a generic picture
with a central emission source surrounded by a roughly spherical
accretion flow. Given the previously mentioned limits on the
millimeter size of \sgra, we could also accomodate models where the
observed 340~GHz emission arises in a small jet component, as the jet
would have to lie within $\sim10r_S$ of the black hole, and would
effectively be a central emission source as seen from a Faraday screen
tens to hundreds of $r_S$ further out. We characterize the radial
density profile, $n(r)$, as a power law,
\begin{equation}
n(r)=n_0(r/r_S)^{-\beta} ,
\label{e-nr}
\end{equation}
where $r_S=2GM_{BH}/c^2$ is the Schwarzschild radius of \sgra\
($10^{12}$~cm for $M_{BH} = 3.5\times10^6M_{\sun}$), and $n_0$ is
the density at this radius. In the case of free-falling gas we have
$\dot M(r)\propto r^p$ with $\beta=3/2-p$, as in
\citet{BlandBegel99}. For spherical accretion \citep{Bondi52} or
Advection-Dominated Accretion Flows \citep[ADAF;][]{NaraYi94} we have
$\beta=3/2$, while for a Convection-Dominated Accretion Flow
\citep[CDAF;][]{QuatGruz00-CDAF}, formally an $\dot M = 0$ limiting case of
convection-frustrated accretion, we have $\beta=1/2$. Intermediate
values are also possible: the best-fit radiatively-inefficient
accretion model in \citet{YuanE03} has $\beta=0.8$, and accretion
flow simulations \citep[e.g.,][]{PenE03} typically produce values
between 1/2 and 1 \citep{Quataert03}. We take the ADAF and CDAF values
as bounds on $\beta$ (i.e., 1/2 to 3/2).

Rather than using a separate parameter to describe the magnetic field
profile, we tie it to the density by assuming equipartition between
magnetic, kinetic, and gravitational energy, as many other modelers
have done \citep[e.g.,][]{Melia92}. For pure hydrogen gas, with the use
of equation (\ref{e-nr}), we obtain
\begin{equation}
B(r)=\sqrt{4\pi c^2 m_H n_o}
\left(\frac{r}{r_S}\right)^{-\left(\beta+1\right)/2} .
\label{e-Br}
\end{equation}
We additionally assume that the magnetic field contains no reversals
along the line of sight and is entirely radial, which should
contribute only a small error unless the field is very nearly
toroidal. The former simplification is a good approximation for
strongly peaked RM vs r profiles (large $\beta$), where only a small
radial range contributes significantly. For smaller $\beta$ and many
field reversals, the effective field will only drop as the square root
of the number of reversals.

In the \sgra\ accretion flow we expect that the electron temperature
($T_e$) will rise to smaller radii, eventually bringing the electrons
to relativistic temperatures ($T_e > 6\times10^9 \mathrm{K} = m_e
c^2/k$) at some radius $r_{in}$.  The RM contribution from
relativistic electrons is suppressed (by as much as
log($\gamma$)/2$\gamma^2$ for Lorentz factor $\gamma$ in the
ultra-relativistic thermal plasma limit; \citealt{QuatGruz00-LP}), so
we approximate this effect by truncating the RM integration at
$r_{in}$ and by treating $r_{in}$ as a variable. From the density
profile, and assuming that gas at $r_{in}$ is in free-fall, we can
determine a mass flux across the $r=r_{in}$ surface
\begin{eqnarray}
\dot M_{in} &=& 4\pi r_{in}^2 m_H n\left(r_{in}\right) v\left(r_{in}\right) \nonumber \\
            &=& 4\pi r_S^2 m_H n_0 c \left(r_{in}/r_S\right)^{3/2-\beta}.
\label{e-mdot}
\end{eqnarray}
This equation does not require that the density profile be followed
down to $r=r_S$, $n_0=n\left(r_S\right)$ is merely a convenient
quatity to normalize the power-law density relation we are assuming
for larger radii. The mass flux at $r_{in}$ ($\dot M_{in}$) can be
taken to be an upper limit on the accretion rate at $r_S$, but the
true rate of accretion onto the black hole could be lower if the
loosely bound plasma falling from $r_{in}$ escapes as a wind or jet.
Substituting equations (\ref{e-nr}), (\ref{e-Br}), and (\ref{e-mdot})
into equation (\ref{e-RM}), and converting $\dot M_{in}$ to units of
\msunyr\ and $r$ to $r_S$, we obtain
\begin{eqnarray}
RM &=& 3.4\times10^{19} 
\left(\frac{M_{BH}}{3.5\times10^6 M_\sun}\right)^{-2} \times \nonumber \\
& & r_{in}^{(6\beta-9)/4} \dot M_{in}^{3/2} 
\int_{r_{in}}^{r_{out}} r^{-\left(3\beta+1\right)/2} dr .
\label{e-dRM}
\end{eqnarray}
Integrating and simplfying yields
\begin{eqnarray}
RM &=& 3.4\times10^{19} 
\left(1-\left(r_{out}/r_{in}\right)^{-\left(3\beta-1\right)/2}\right) 
	\times \nonumber \\
& & \left(\frac{M_{BH}}{3.5\times10^6 M_\sun}\right)^{-2}
	\left(\frac{2}{3\beta-1}\right) r_{in}^{7/4}\dot M_{in}^{3/2} .
\label{e-RM-mdot}
\end{eqnarray}
To obtain an RM given $\beta$ and $\dot M_{in}$ we must also choose
$r_{in}$ and $r_{out}$. The inner radius will vary by model, but it is
typically around $300r_S$ \citep[e.g.,][]{YuanE03}. For these
calculations we consider values of $r_{in}$ from 300 to $3r_S$ in
order to account for variations among models and to allow for the
possibility that the electrons do not become highly relativistic
interior to $r_{in}$, in which case the RM would not be strongly
suppressed. The outer radius depends on the coherence of the radial
field. We examine two cases: a fully coherent field
($r_{out}\approx\infty$), and a field that persists for a factor of
three in radius from $r_{in}$.

\begin{figure}
\plotone{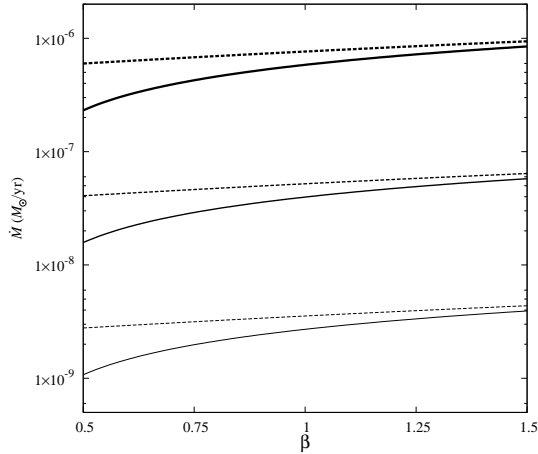}
\caption{Accretion rate limits imposed by the rotation measure limit
of $7\times10^5$~\rmm\ as a function of the density power law, given
the accretion model described in \S~\ref{s-mdot}. The accretion rate
plotted here is measured at the radius where the electrons become
relativistic, $r_{in}$; extrapolation to the black hole event horizon
is discussed in \S~\ref{s-mdot}. Two sets of curves are plotted (see
eq. [\ref{e-RM-mdot}]): three for a magnetic field that is coherent to
large radius (solid lines) and three for a field that is coherent over
a smaller range ($r_{out}/r_{in}=3$, dashed lines). Within each set,
the thickest line is $r_{in} = 300r_S$, then $30r_S$, and finally
$3r_S$.}
\label{f-mdot}
\end{figure}

Figure~\ref{f-mdot} shows the accretion rate limits imposed by our RM
limit of $7\times10^5$~\rmm, based on the model described above. From
the two choices of $r_{out}$ we see that the effect of the magnetic
field coherence is larger at small $\beta$. As mentioned before, for
steep density profiles (large $\beta$) we expect that only a small
range in radius around $r_{in}$ contributes to the RM, making the
inferred accretion rate limit insensitive to the field coherence
length. If we assume that the density profile follows equation
(\ref{e-nr}) down to $r=r_S$, our model imposes accretion rate limits
that are a factor of $\dot M\left(r_S\right) / \dot M_{in} =
(r_{in}/r_S)^{\beta-3/2}$ lower than those in Figure~\ref{f-mdot}, but
the transition to supersonic flow makes this density extrapolation
uncertain. However, in cases like the basic ADAF model
\citep{NaraYi95} where the electron temperature ceases to rise at
small radii and the electrons are only marginally relativistic,
integration to smaller radii (the lower sets of curves) may set more
relevant (and lower) accretion rate limits. In fact, taking $\beta =
3/2$ and $r_{in} = 30r_S$, we roughly have the ADAF/Bondi model used in
\citet{QuatGruz00-LP}, and reproduce their $\dot M$ limit of
$10^{-7}$~\msunyr. The high and low-$\beta$ limits are similar, but the
field coherence is a larger concern for shallow profiles. Since the
prototype for a low-$\beta$ model is a highly convective flow we may
expect a tangled field, but in this case the accretion rate limit
(proportional to $B^{-2/3}$) will increase only as $\dot M \propto
N^{1/3}$ for $N$ field reversals. In summary, the figure shows that
for any choice of density profile, the maximum allowed accretion rate
is $10^{-6}$~\msunyr, and may be much lower. This is an order of
magnitude below the gas capture rate of $10^{-5}$~\msunyr\ inferred
from X-ray observations \citep{BaganoffE03,YuanE03} and from
simulations of stellar winds in the Galactic Center
\citep[e.g.,][]{Quataert04,CuadraE05}. It is therefore likely that
there is substantial mass lost between the gas capture at
$r\sim10^5r_S$ and the event horizon.

Finally, this model of the accretion flow can be used to examine the
proposed $-3.5\times10^6$~\rmm\ RM from May 26
(\S~\ref{s-discRM}). This RM would require a change of more than
$2\times10^6$~\rmm\ between consecutive nights. This is very large
compared to the RM changes implied by other position angle changes
(again assuming that the source polarization remains constant). 
Based on the four other nights
with strong polarization detections, all of which have position angles
near to 140\degr, the peak-to-peak $\chi$ change corresponds to an RM
change of $1.5\times10^5$~\rmm\ and the rms variation is only
$5\times10^4$~\rmm. The largest change on similar (day to week)
timescales observed at 230~GHz is $3\times10^5$~\rmm\ \citep[between
2003 December 27 and 2004 January 5;][]{BowerE05}. The longer
timescale 230~GHz position angle changes and the difference between
our position angles and the \citet{AitkenE00} 350~GHz position angle
(reinterpreted as described in \S~\ref{s-LPvar} or otherwise) also
correspond to RM changes of a few$\times10^5$~\rmm. We expect that
these variations are not more than order unity fractional RM changes,
so they are all quite consistent with our inferred
$-5\times10^5$~\rmm\ quiescent RM from \S~\ref{s-discRM}. The May 26
RM would then correspond to a factor of 7 increase in the density or
line of sight magnetic field. Such a change is difficult to accomplish
with any density profile, but is particularly difficult for small
$\beta$ where the entire line of sight contributes significantly to
the RM. If the fluctuation is real it suggests a steep density
profile, as the associated density/field change should not be extended
over decades of radius. Unless such an event is observed again in
future observations, the more likely interpretations appear to be that
the position angle change from May 25 represents a RM fluctuation of
$4\times10^5$~\rmm\ observed between consecutive nights or a
transient change in the source polarization, and the May
26 difference in the USB and LSB position angles is merely a $2\sigma$
measurement noise event.

\subsection{Linear Polarization and Variability}
\label{s-LPvar}
Our 340~GHz observations show a typical polarization fraction of
6.4\%, with a range of 2.3$-$8.5\%, and an rms variation of
2.0\%. This is comparable to the $\sim$7.5\% mean, $4.6-13.6$\% range,
and 2.2\% rms measured at 230~GHz by \citet{BowerE03,BowerE05}. The
range of observed polarization is lower at 340~GHz than it is at
230~GHz, and the mean is slightly lower as well. It is difficult to
explain a lower observed polarization fraction (and comparable
variability) at higher frequencies with beam depolarization models
\citep{Tribble91}, as Faraday rotation and the resulting dispersion in
polarization directions decreases with increasing frequency. If the
polarization fraction decrease is intrinsic to the source and not
generated in the propagation medium, it suggests that the magnetic
field becomes increasingly disordered at smaller radii, as these
observations should probe slightly smaller radii than the 230~GHz
data. But across only 0.2 decades in frequency we expect little change
in intrinsic polarization, so the difference, if present, may be best
explained by time variability in the source polarization. To resolve this question, simultaneous
or nearly-coincident polarimetry at multiple frequencies with
interferometer resolution is clearly desirable.

\citet{BowerE05} used the apparent stability of the 230~GHz
polarization fraction to argue that the observed variations in the
230~GHz polarization position angle were more likely to be the result
of changes in the rotation measure than due to intrinsic source
changes. While our results do not refute this conclusion, they
demonstrate that the polarization fraction is not stable, even over a
single night. Note that two substantial excursions in the 230~GHz
polarization fraction, one of which is labeled an ``outlier'' in
\citet{BowerE05}, probably represent real variations similar to those
seen here, but have lower significance because of the poorer
sensitivity of their instrument.

The polarization fraction presented here is considerably lower than
those measured in 1999 by \citet{AitkenE00}: $13^{+10}_{-4}$\% and
$22^{+25}_{-9}$\% at 350 and 400~GHz, respectively. However, to
determine the flux density of \sgra\ \citet{AitkenE00} had to correct
for the contamination from dust and free-free emission in their large
primary beam (14\arcsec$-$12\farcs5 at the highest frequencies), and
it is possible that they over-corrected for the dust emission, which
would make the polarized component appear to be a larger fraction of
the total flux density of \sgra.  There is some support for this
possibility from their measured flux densities: \sgra\ was found to be
only 2.3 and 1~Jy at 350 and 400~GHz, while our data (see
Table~\ref{t-pol}) and previous measurements between 300 and 400~GHz
have found higher values of $2.6-3.8$~Jy
\citep{ZylkaE95,SerabynE97,PiercePriceE00}. If we assume that their
350~GHz data are well calibrated (the 400~GHz calibration is more
uncertain) and assume our 3.3~Jy flux density for \sgra, we can
re-derive the intrinsic polarization of \sgra\ using their Stokes Q
and U decomposition method, and find a polarization of 9\% at
158\degr. The polarization fraction drops further as the assumed flux
density for \sgra\ is increased, reaching 7.6\% for 3.8~Jy. These
values are within the polarization fraction variations we observe; one
might expect that well calibrated 400~GHz measurements could be
interpreted similarly and that the polarization fraction need not rise
steeply to high frequencies. In arriving at a flux density of 2.3~Jy
for \sgra, \citet{AitkenE00} estimated the dust emission in their
central pixel from the average of the surrounding pixels, so by
increasing the contribution from \sgra\ we are also suggesting that
there is a deficit of dust emission in the central 14\arcsec at
350~GHz. Unfortunately, our observations are poorly sampled at short
spacings, but the available visibilities shortward of 20~\kl\ show
little excess over the point source flux density, consistent with such
a central hole in the dust emission. The existence of this hole
requires further confirmation, as could be achieved through
simultaneous single-aperture and interferometer observations; our
circumstantial evidence could be equally well explained if \sgra\ had
a higher polarization fraction and lower flux density in 1999 (at the
time of the \citet{AitkenE00} measurement) and if the emission in the
central 30\arcsec\ is distributed smoothly on scales smaller than
10\arcsec.

We observe variability on inter-night and intra-day intervals, in both
the polarization and total intensity. The single-night flux densities
we measure fall within the range of previous observations, and the rms
variation of 0.3~Jy, or 10\%, matches the recent results of
\citet{MauerhanE05} at 100~GHz. Within nights, the Stokes I light
curves in Figure~\ref{f-plc} show unambiguous variations on timescales
of hours, reminiscent of those seen at 100 and 140~GHz by
\citet{MiyazakiE04} and \citet{MauerhanE05}. This is slower than the
variations seen in the near-infrared and X-ray
\citep[e.g.,][]{BaganoffE01,GenzelE03}, which seem to vary on hour
timescales, with some features requiring only minutes. These slow
changes suggest that opacity is obscuring our view of the very inner
regions of the accretion flow, regions unobscured at NIR/X-ray
wavelengths, even at 340~GHz. At slightly higher frequencies the inner
flow may become visible, although many estimates of the optically-thin
transition frequency place it at or above 1~THz, a frequency that is
difficult to access from the ground. It should be possible to search
for the transition to optically thin emission using the change in the
variability timescale; the more frequently proposed technique of
looking for the turnover in the spectrum relies on precise flux
density calibration at high frequencies, which is problematic because
of contaminating emission in single-aperture beams and lack of
unresolved calibrators in interferometers. A few instruments may be
able to make these difficult observations before ALMA: the SMA, or
perhaps SCUBA \citep{SCUBA} on the JCMT at 650~GHz, and SHARC~II
\citep{SHARCII} on the CSO at 650 or 850~GHz.

The intra-day variations in the linear polarization shown in
Figure~\ref{f-plc} are the first linear polarization changes observed
on intervals of hours rather than days. The three nights with
time-resolved polarization measurements do not demonstrate a clear
relationship between Stokes I and the polarization. For example, May
25 shows a very strong flare in I with $m$ very close to our
average values, followed in the second half of the track by a lower I
and a below average $m$. July 5 has the highest $m$ of our six
nights, along with 20\% modulation in I, but the polarization fraction
is not modulated significantly with the total intensity. Finally, on
July 6 we see below average $m$ in a period of high I and above
average $m$ with low I, the inverse of the relationship seen on
May 25. That the polarization fraction may vary in multiple ways
during flares in the total intensity could suggest that there are
multiple mechanisms (of varying polarization) responsible for the
submillimeter Stokes I variability, or that the I and $m$ changes
are not closely related. Diverse flare mechanisms could be expected to
show different spectra at shorter wavelengths, so simultaneous
infrared and X-ray data may be useful. However, based on the
infrequency of infrared and X-ray flares \citep{EckartE04} and the
lack of coincident activity in these bands during the SMA observations
on July 6 and 7 \citep{EckartE05}, it seems that the small changes we
observe in the submillimeter are often imperceptible at shorter
wavelengths. Therefore, the best way to determine whether the
polarization changes are internal or external may be to increase the
time resolution in the polarization light curves. In these data we
observe $m$ changes on the shortest interval we can measure,
around three hours (on July 6). This is close to the variability
timescale observed in the total intensity, which suggests that given
better time resolution we may see that the I and $m$ changes have
similar temporal characteristics and therefore arise from the same
processes.

The $m$ and $\chi$ curves seem to show more coordinated behavior
than the total intensity and polarization do. Of the seven sub-night
intervals plotted in Figure~\ref{f-plc}, five show position angles
close to the observed quiescent $\chi$ of 140\degr. Only in the two
intervals with the lowest polarization, on May 25 and July 6, does
$\chi$ deviate from this value, and if the deviations are caused by RM
changes then both would represent increases in the RM. None of the
intervals provide evidence for a RM through inter-sideband $\chi$
differences, but the largest $\chi$ change between intervals,
$-20.7^\circ\pm3.8^\circ$ on July 6, only requires a RM decrease of
$5\times10^5$~\rmm, still below our detection limits.  Here again we
face the question of whether the source polarization or an external
process is responsible for the variability we see. It is possible to
explain the $\chi$ changes with a two-component source, where the
dominant polarization component is polarized close to the quiescent
polarization direction and variable in amplitude while the weaker
component causes the polarization to deviate from 140\degr\ when the
dominant component weakens. In this case we would expect to see a
correlation between the polarization fraction and the position angle,
something that is not excluded by our data. Such a source model is
naturally identified with emission from a core and jet.  A second
model uses a turbulent plasma screen, in addition to the screen
responsible for the putative mean RM (suggested by the difference in
position angle between 230 and 340~GHz), to partially beam depolarize
the emission. The fact that $\chi$ seems to faithfully return to
140\degr\ implies that the source, or the source plus a stable RM
component, is separated from the changes that cause the depolarization
and position angle change. With better time resolution and better
sensitivity to RM it should be possible to distinguish between these
models.

\section{Conclusions}
\label{s-concl}
Using the Submillimeter Array, outfitted with polarization conversion
hardware (Marrone, in preparation), we have made sensitive
measurements of the polarization of \sgra\ at 340~GHz with angular
resolution sufficient to separate the source from the surrounding
contaminating emisssion. Our increased sensitivity has allowed us to
make unequivocal measurements of the variability of the linear
polarization of this source, in both position angle and polarization
fraction. This is the first reliable detection of variation in the
linear polarization fraction. Moreover, we have made the first
detection of linear polarization changes within a night. These changes
do not show an obvious correlation with the observed changes in the
total intensity, possibly because of the coarse time resolution
available at our sensitivity limits. The polarization variations occur
on the shortest intervals we sample, around 3 hours, which is
comparable to the modulation time observed in the total intensity here
and in
\citet{MauerhanE05} at 100~GHz. It is not clear from these data
whether the polarization variability can be best explained by changes
in the source emission or by changes in an external Faraday screen,
but polarization light curves with better time resolution should
clarify the issue. The observed polarization fraction at 340~GHz is
comparable to, and perhaps lower than, that observed at 230~GHz. This
contradicts the polarization spectrum measured from 150 to 400~GHz by
\citet{AitkenE00}, but we show that their polarization fraction at
350~GHz can be brought into agreement with ours through changes in
their correction for dust emission. Whether or not the polarization
fraction rises steeply to high frequency, as predicted by synchrotron
optical depth explanations of the early polarization results
\citep{Agol00,MeliaLiuCoker00}, is no longer clear, but this question
should be resolved by future submillimeter polarimetry at 650~GHz.

We have also measured the circular polarization of this source to be
less than 0.5\% for a time-stable component, and do not detect CP at a
slightly higher level in individual nights. This limit contradicts the
predictions of the turbulence-driven polarization conversion model of
\citet{Beckert03}, which was designed to match the \citet{AitkenE00}
linear polarization results, but can be matched to an earlier version
of the model \citep{BeckertFalcke02} where the CP originated in a
fully turbulent jet.

By comparing the position angles in the two sidebands, we place new
upper limits on the RM allowed for this source. In single nights we
obtain one-sigma upper limits of less than $10^6$~\rmm\, with our
lowest limit of $7\times10^5$~\rmm\ coming on July 5. This is
comparable to the lowest limit obtained in any other polarimetric
observations of this source and well below the single-night limits of
other interferometers. We can use a model accretion flow (with energy
equipartition), parameterized only by the density power-law slope and
the radius at which the electrons become relativistic, to convert this
RM to a mass accretion rate limit, and find that for any density slope
\sgra\ is accreting at least an order of magnitude less matter than it
should gravitationally capture based on X-ray measurements
\citep{BaganoffE03}, and may be accreting much less if the density
profile is shallow. This result agrees with earlier interpretations of
polarization detections. We note that the position angle at 340~GHz
seems to show a persistent stable state, much like that observed at
230~GHz \citep{BowerE05}, and we combine these two values to infer a
stable ``quiescent'' RM of $-5\times10^{-5}$~\rmm. This value is just
below the detection limit of our observations. The possible proximity
of the RM to the detection threshold, the need for more time-resolved
polarimetry, and the potential for coordinated observations with other
wavelengths suggests that expanded SMA capabilities may contribute
considerably more to this study.

\acknowledgements
The authors thank the entire SMA team for their contributions to the
array and to the new polarimetry system. In particular, we acknowledge
the enormous contribution of K. Young for his work on the real-time
software changes essential to these observations. We thank R. Narayan,
E. Quataert, and G. Bower for useful discussions, and J. Greene for
discussions and her help in developing the prototypes of the
polarimetry system. DPM was supported by an NSF Graduate Research
Fellowship. We thank an anonymous referee for a thorough reading and
helpful comments. Finally, we extend our gratitude to the Hawaiian
people, who allow us the privilege of observing from atop the sacred
mountain of Mauna Kea.

\end{document}